\documentclass[12pt]{article}
\newcommand{\sect}[1]{\setcounter{equation}{0}\section{#1}\indent}

\renewcommand{\thefootnote}{\fnsymbol{footnote}}
\newcommand{\EQ}{\begin{equation}}
\newcommand{\EN}{\end{equation}}
\newcommand{\bea}{\begin{eqnarray}}
\newcommand{\ena}{\end{eqnarray}}
\newcommand{\vs}[1]{\vspace{#1 mm}}

\renewcommand{\a}{\alpha}

\newcommand{\uda}{\nearrow \kern-1em \searrow}

\newcommand{\NP}[3]{Nucl.~Phys.~{\bf #1} (#2) #3}

\newcommand{\PR}[3]{Phys.~Rev.~{\bf #1} (#2) #3}

\newcommand{\hepth}[1]{[arXiv:hep-th/#1]}

\newcommand{\La}{\Lambda}

\newcommand{\beq}{\begin{equation}} 
\newcommand{\eeq}{\end{equation}}
\newcommand{\beqs}{\begin{eqnarray}} 
\newcommand{\eeqs}{\end{eqnarray}}
\newcommand{\tq}{\tilde{Q}}
\newcommand{\tb}{\tilde{b}}
\newcommand{\tB}{\tilde{B}}
\newcommand{\tD}{\tilde{\Delta}}
\newcommand{\tQ}{\tilde{Q}}
\newcommand{\Qt}{\tilde{Q}}

\newcommand{\lit}[3]{#1, ``{\it #2}'', #3.}
\newcommand{\lits}[2]{#1, #2.}

\begin{document}

\begin{titlepage}
\setcounter{page}{0}
\begin{flushright}
EPHOU 02-009\\
December 2002\\
\end{flushright}

\vs{6}
\begin{center}
{\bf{{\Large  Mean-field Approach to the Derivation of Baryon Superpotential from Matrix Model}}}

\vs{6}
{\large
Hisao Suzuki}\\
\vs{6}
{\em Department of Physics, \\
Hokkaido
University \\  Sapporo, Hokkaido 060 Japan\\
hsuzuki@phys.sci.hokudai.ac.jp} \\
\end{center}
\vs{6}

\centerline {{\bf {Abstract}}}
We discuss how to obtain the superpotential of the baryons and mesons for $SU(N)$ gauge theories with $N$ flavour 
t
\newpage
\renewcommand{\thefootnote}{\arabic{footnote}}
\setcounter{footnote}{0}
\end{titlepage}
\sect{Introduction}
Recently, Dijkgraaf and Vafa~\cite{DV1,DV2,DV3} proposed a general prescription for computing effective glueball superpotentials via planar diagrams of matrix
theories. Several tests have been achieved for this conjecture~\cite{Chekhov}-~\cite{IM2} for the fields in adjoint
representations.  A recent paper~\cite{DGLVZ,CDSW} proved a correspondence for matter fields in the adjoint representation. A natural extension of
the construction is to the theories with fields in the fundamental representations. 
 The gauge theory effective potential for glueball superfield $S$ is given by
\beq
    W_{DV}(S, \Lambda, \lambda_a) = N_c S (-\log(S/\Lambda^3) +1) + 
       N_c \frac{\partial {\cal F}_{\chi=2}(S, \lambda_a)}{\partial S} +
        {\cal F}_{\chi=1}(S, \lambda_a).
\eeq
where ${\cal F}_{\chi=2}$ is the contribution from adjoint fields and ${\cal F}_{\chi=1}(S, \lambda_a)$ is the contribution from the fields in the
fundamental representations. The programm was successfully used to compare matrix model results with known superpotentials of meson fields~\cite{ACFH}-\cite{Ohta}.

The advantage the matrix theory should be  that we can prove some basic hypothesis or conjectures which cannot be provided by the standard
field theoretical methods. In fact, a derivation of the non-renormalization theorem was performed in ~\cite{Tachikawa2}.  

 One of the
biggest problems is the inclusion of the baryonic fields. Matrix theory may be useful to get insights for the baryons. This problem was treated in a
recent papers~\cite{ACFH2,BRT}. The paper \cite{ACFH2} showed that the first order calculation in the perturbative technique agrees with the field theoretical result. In
\cite{BRT}, they showed that for
$SU(2)$ gauge fields with $2$ flavor models, we can obtain superpotential for baryonic fields. In this model, the baryons can be expressed by bilinear
fields and the matrix integration can be performed. However, for $SU(N)$, $N>2$, the baryons are composed of more than trilinear fields so that we
cannot easily  perform the matrix integrations. 

In this paper, we consider the theories with
$SU(N)$ gauge fields with
$N$ flavour.  The quantum moduli space is conjectured  to be parametrized by meson fields and baryons by~\cite{Seiberg2,IS,ILS}
\bea
{\rm det}{M}-B\tilde{B}=\La^{2N}.
\ena
Our aim of this paper is the derivation of this constraint from the matrix model. Since baryonic perturbation is complicated, we will use the mean-field method to
estimate the superpotential, which is related to the free energy of the matrix model. It will turn out that the self-consistency equation for the mean field depends on
$N$. We will therefore assume a planar limit of the equation. With this assumption, we will show that we can get a potential which almost agrees with the expected
results. Before doing this, we will first treat the field theoretical result which should be compared to the matrix theory result.
\sect{Effective Superpotentials for $N_c =N_f$}
We will first argue how to obtain effective superpotential from the field theory. We consider $SU(N)$ gauge theories with $N$ quark flavors $Q^i_a$ and
$\tilde{Q}_i^a$. We diagonalize the mass matrix and consider the tree level superpotential
\bea
W_{\rm tree}(Q,\tq)= m_i\tilde{Q}_i^aQ^i_a +b\det Q +\tilde{b}\det\tq, 
\ena
which can be written in terms of meson operators and baryon operators
\bea
M_i^j=Q_i^a\tq^j_a,\qquad B=\det Q, \qquad \tilde{B}=\det\tq
\ena
as
\bea
W_{\rm tree}(X_i,B,\tilde{B})= m_iX_i+bB +\tilde{b}\tilde{B},
\ena
where we have defined $X_i=M_i^i$.
In order to obtain the non perturbative part of the effective superpotential, we  start with
the Veneziano-Yankielowicz superpotential~\cite{VY}.
\bea
W=NS-S\ln\frac{S^N}{\tilde{\La}^{3N}}.
\ena
For $N$-flavour, we assume the matching relation ${\tilde{\La}}^{3N}=m_1m_2\cdots m_N\La^{2N}$.
When we integrated in $X_i$ fields as
$X_i=\partial_{m_i}W$, we get the following superpotential
\bea
W=S \ln\frac{\La^{2N}}{X_1X_2\cdots X_N} + m_iX^i,
\ena
which can be promoted to
\bea
W=S \ln\frac{\La^{2N}}{{\rm{det}}M}+m_iX^i
\ena
If we include the baryon fields, it is natural to expect that
the effective potential is given by 
\bea
W=S \ln\frac{\La^{2N}}{-B\tB+{\mathrm{det}M}}+bB+\tb\tB +m_iX^i.
\label{effective} 
\ena
The equation $\partial_SW=0$ implies the constraint 
\bea
{\rm det}{M}-B\tilde{B}=\La^{2N},
\ena
which is a desired result~\cite{Seiberg2,IS,ILS}. We assume that this potential is exact although the proof of the
non-renormalization theorem of baryoionic perturbation may be very subtle.

To compare this result with the superpotential from the matrix models, we are going to integrated out the fields $B,\tB$ and $M_i^j$.
Let us first denote 
\bea
\Delta={\mathrm{det}}M-B\tB
\ena
The equation $\partial_BW=0, \partial_{\tB} W=0$ implies
\bea
B=-\frac{\Delta \tb}{S},  \tB=-\frac{\Delta b}{S}.
\ena
The variation with respect to $M_i^j$ leads to
\bea
M_i^j&=&0,  (i\neq j) \nonumber\\
m_iX_i&=&(\frac{\Delta}{S}m_1m_2\cdots m_N)^{\frac{1}{N-1}}.
\ena
From these relations, we find that $\Delta$ satisfies an equation
\bea
\Delta=X_1X_2\cdots X_N-B\tB =(\frac{\Delta}{S})^{\frac{N}{N-1}}-(\frac{\Delta}{S})^2b\tb.
\ena
when $b=0$, this equation has a solution $\Delta=S^N/m_1m_2\cdots m_N$. Therefore, we define a new
variable by $\Delta=S^N/m_1m_2\cdots m_N\tD$. Then the equation for $\tD$ is given by
\bea
\tD=(1+\a\tD)^{N-1},
\label{valueofdtilder}
\ena
where $\a$ is defined as 
\bea
\a=\frac{S^{N-2}b\tb}{m_1m_2\cdots m_N}.
\label{valueofalpha}
\ena
Substituting these variables in (\ref{effective}), we find that the effective superpotential can be written in the form
\bea
W=NS(1-\ln S/\La^3) +\cal{F},
\ena
where $\cal{F}$ is 
\bea
{\cal{F}}=S\ln\frac{m_1m_2\cdots m_N}{\La^N}-S \ln\tD+(N-2)\a S\tD.
\label{free energy of superpotential}
\ena
This is the final form of the superpotential which we want to compare in the next section. Note also that the superpotential (\ref{free energy of superpotential}) agrees with
the one obtained in ref~\cite{ACFH2} although  the derivation seems different.
\sect{Mean-field Approximation for Matrix Model}
\subsection{Mean Field Approach}
Before discussing the model for baryonic perturbation, we will apply the mean-field
method to the known models. The aim of doing this is  to get a support for the validity of the mean-field method. 
The tree level superpotential  we treat here is obtained by the deformation from $N{=}2$ SQCD
theory with fundamental flavors $Q_i^a$  and $\Qt_i^a$ by adding a mass $M$ for the adjoint scalar
$\Phi_a^b$ in $N{=}2$ vector multiplet
\bea
W_{\rm tree}(\Phi,Q,\Qt)=\frac{1}{2}M{\mathrm{Tr}} \Phi^2
+ g\Qt^a_i \Phi_a^b Q^i_b
+ m \Qt_i^a Q_a^i,
\label{Wtree}
\ena
where we have set the mass of the fundamental flavors to be $m_1=m_2=\dots=m_{N_f}=m$ for simplicity.
We consider the following free energy 
\bea
e^{-\frac{N}{S}{\cal{F}}}=(\frac{N\La}{S})^{N_f}\int d\Phi dQ d\tq e^{-\frac{N}{S}W_{\rm tree}(\Phi,Q,\Qt)},
\ena
Since the integral for the adjoint fields $\Phi_a^b$ is Gaussian, we can integrate out the fields and obtain
\bea
e^{-\frac{N}{S}{\cal{F}}}=(\frac{N\La}{S})^{N_f}\int d\Phi dQ d\tq e^{-\frac{N}{S}[m \Qt_i^a Q_a^i-\frac{g^2}{2M}\Qt_i^a Q_a^j\Qt_j^b
Q_b^i]},
\ena
We re-scale the variables as
\bea
Q_i^a \rightarrow (\frac{N}{S}m)^{-1/2}Q_i^a,\qquad \tq^i_a\rightarrow (\frac{N}{S}m)^{-1/2}\tq^i_a,
\ena
to get
\bea
e^{-\frac{N}{S}{\cal{F}}}=(\frac{\La}{m})^{NN_f}\int d\Phi dQ d\tq e^{-W},
\ena
where
\bea
 W=\Qt_i^a Q_a^i-\frac{\beta S}{2N}\Qt_i^a Q_a^j\Qt_j^bQ_b^i,\qquad \beta=\frac{g^2}{mM^2}.
\ena
We will write the correlation function $<\Qt_i^a Q_b^j>$ as
\bea
<\Qt_i^a Q_b^j>=\Delta^\prime\delta_i^j\delta^a_b,
\label{anzatz}
\ena
Then the above tree level potential$W$ can be written in the form
\bea
W=&{}&\Qt_i^a Q_a^i-\beta S\Delta^\prime \Qt_i^a Q_a^i+NN_f\frac{\beta S}{2}\Delta^{\prime2} \nonumber\\
&{}&-\frac{\beta
S}{2N} \sum_{i,j=1}^{N_f}\{\sum_{a=1}^N(\Qt_i^a Q_a^j-\Delta^\prime\delta_i^j)][\sum_{b=1}^N(\Qt_j^b
Q_b^i-\Delta^\prime\delta^i_j)]\}.
\ena
As a mean field theory, we neglect the contribution of the last term. Namely, we consider the
following free energy:
\bea
e^{-\frac{N}{S}{\cal{F}}}=(\frac{\La}{m})^{NN_f}\int d\Phi dQ d\tq e^{-[\Qt_i^a Q_a^i-\beta S\Delta^\prime \Qt_i^a
Q_a^i+NN_f\frac{\beta S}{2}\Delta^{\prime2} ]},
\label{ap}
\ena
We can compute the correlation function  $<\Qt_i^a Q_b^j>=\Delta^\prime\delta_i^j\delta^a_b$
and get a consistency equation for $\Delta^\prime$:
\bea
\frac{1}{1-\beta S \Delta^\prime}=\Delta^\prime.
\label{consistency}
\ena
This equation can be easily solved by 
\bea
\Delta^\prime=\frac{1-\sqrt{1-4\beta S}}{2\beta S}=\frac{2}{1+\sqrt{1-4\beta S}}.
\ena
Performing the Gaussian integral in (\ref{ap}), we get a free energy
\bea
{\cal{F}}&=&N_fS[\ln\frac{\La}{m}-\ln\Delta^\prime+\frac{\beta S}{2}\Delta^{\prime2})]\nonumber\\
&=&N_fS[\ln\frac{\La}{m}-\frac{1}{2}-\frac{1}{4\beta S}(1-\sqrt{1-4\beta S})+\ln\frac{1+\sqrt{1-4\beta S}}{2}]
\ena
which is exactly the result obtained by summing up the ladder diagrams\cite{ACFH}.

In general, we may have $N$ dependence in the self-consistency equation (\ref{consistency}). Then, we should take the planar
limit where
$N$ dependence disappears. Then we expect that the result reflects the contribution from the planar diagrams.
\subsection{Effective Superpotential for Baryons from Matrix Theory}
Let us apply the mean-field technique to the baryonic perturbation.
We consider the free energy
\bea
e^{-\frac{N}{S}{\cal{F}}}=(\frac{N\La}{S})^N\int \prod_{i,a=1}^NdQ_i^a d\tq^i_ae^{-\frac{N}{S}W_{tree}(Q,\tq)},
\ena
which is in our case,
\bea
&{}&e^{-\frac{N}{S}{\cal{F}}} \nonumber\\
&=&(\frac{N\La}{S})^N\int \prod_{i,a=1}^NdQ_i^ad\tq^i_ae^{-\frac{N}{S}(m_i\tq_i^aQ_a^i+b\epsilon^{a_1a_2\cdots a_N}Q_{a_1}^1Q_{a_2}^2
\cdots Q_{a_N}^N+\tb\epsilon_{b_1b_2\cdots b_N}\tQ_1^{b_1}\tQ_2^{b_2}\cdots\tQ_N^{b_N})}, \quad\quad
\ena
The superpotential can be derived by the free energy $\cal{F}$ as
\bea
W=NS(1-\ln \frac{S}{\La^3})+{\cal{F}}.
\ena
The first term of the superpotential is the Veneziano-Yankielowicz superpotential which can be obtained either from the volume of the gauge groups~\cite{DGLVZ}
or Gaussian integral measures~\cite{CDSW}.
We will first re-scale the variables 
\bea
Q_i^a \rightarrow (\frac{N}{S}m_i)^{-1/2}Q_i^a,\qquad \tq^i_a\rightarrow (\frac{N}{S}m_i)^{-1/2}\tq^i_a.
\ena
Then the matrix integral can be written as 
\bea
&{}&e^{-\frac{N}{S}{\cal{F}}} \nonumber\\
&=&(\frac{\La^N}{m_1m_2\cdots
m_n})^N\int \prod_{i,a=1}^NdQ_i^a
d\tq^i_ae^{-W}.
\ena
where
\bea
&{}&W=\tq_i^aQ_a^i\nonumber\\
&{}&-[(\frac{S}{N})^{N-2}\frac{1}{m_1m_2\cdots m_n}]^{1/2}(b\epsilon^{a_1a_2\cdots
a_N}Q_{a_1}^1Q_{a_2}^2
\cdots Q_{a_N}^N+\tb\epsilon_{b_1b_2\cdots b_N}\tQ_1^{b_1}\tQ_2^{b_2}\cdots\tQ_N^{b_N})\quad\quad
\ena
Integrating out the fields $Q_a^N$ and $\tq_N^a$, we get 
\bea
e^{-\frac{N}{S}{\cal{F}}} =(\frac{\La^N}{m_1m_2\cdots
m_n})^N\int\prod_{i=1}^{N-1}\prod_{a=1}^NdQ_i^ad\tq^i_ae^{-\tilde{W}}
\label{master}
\ena
where $\tilde{W}$ is given by
\bea
\tilde{W}=-\tq_i^aQ_a^i+\frac{\a }{N^{N-2}}\epsilon^{a_1a_2\cdots
a_N}\epsilon_{b_1b_2\cdots a_N}Q_{a_1}^1Q_{a_2}^2
\cdots Q_{a_{N-1}}^{N-1}\tQ_1^{b_1}\tQ_2^{b_2}\cdots\tQ_{N-1}^{b_{N-1}},
\ena
and the parameter $\a$ is defined in (\ref{valueofalpha}). 
To estimate the free energy, we apply the mean-field approximation. We put the correlation
functions
$<\epsilon^{aa_1a_2\cdots a_{N-1}a_N}\epsilon_{b_1b_2\cdots b_{N-1}a_n}\prod_{k\neq
i}^{N-1}Q_{a_i}^k\tQ^{b_i}_k>$ in the form
\bea
<\epsilon^{aa_1a_2\cdots a_{N-1}a_N}\epsilon_{b_1b_2\cdots b_{N-1}a_n}\prod_{k\neq
i}^{N-1}Q_{a_i}^k\tQ^{b_i}_k>=(N-1)N^{N-2}\Delta^\prime\delta_{a_i}^{b_i}.
\label{gap}
\ena
Then we can rewrite $\tilde{W}$ as
\bea
&{}&{\tilde{W}}=-\tq_i^aQ_a^i+\a \Delta^\prime\tq_i^aQ^i_a-\a \Delta^\prime<Q^i_a\tq_i^a>
\nonumber\\
&{}&+\sum_{i=1}^{N-1}\frac{\a \epsilon^{a_1a_2\cdots a_N}\epsilon_{b_1b_2\cdots
a_N}}{(N-1)N^{N-2}}\prod_{k\neq i}^{N-1}Q_{a_i}^k\tQ^{b_i}_k<Q^i_{a_i}\tq_i^{b_i}>\nonumber\\
&+&\a \sum_{i=1}^{N-1}(\frac{\epsilon^{a_1a_2\cdots a_N}\epsilon_{b_1b_2\cdots
a_N}}{(N-1)N^{N-2}}\prod_{k\neq i}^{N-1}Q_{a_i}^k\tQ^{b_i}_k-\Delta^\prime\delta^{a_i}_{b_i})(Q_{a_i}^i\tq^{b_i}_i-<Q_{a_i}^i\tq^{b_i}_i>) 
\label{full}
\ena
As a mean field evaluation, we neglect the contribution of the last term of equation (\ref{full}). We also omit the contribution of the term 
\bea
\sum_{i=1}^{N-1}\frac{\a \epsilon^{a_1a_2\cdots a_N}\epsilon_{b_1b_2\cdots
a_N}}{(N-1)N^{N-2}}\prod_{k\neq i}^{N-1}Q_{a_i}^k\tQ^{b_i}_k<Q^i_{a_i}\tq_i^{b_i}>.
\label{negrect}
\ena
Note that the omitting of this term is not completely valid for $N=2$ and $N=3$. For $N=2$, this term is a constant and should be included in the free
energy. This case, the integral in (\ref{master}) can be performed without any approximation. We can easily see that the resulting expression of free
energy agrees with the field theoretical results(\ref{free energy of superpotential}) for $N=2$. For $N=3$, (\ref{negrect}) is bilinear in fields and
we should prepare more generic ansatz for the correlator
\bea
<\epsilon^{aa_1a_2\cdots a_{N-1}a_N}\epsilon_{b_1b_2\cdots b_{N-1}a_n}\prod_{k\neq
i}^{N-1}Q_{a_i}^k\prod_{k\neq j}\tQ^{b_i}_k>=(N-1)N^{N-2}\Delta^{\prime b_j,i}_{a_i,j},
\label{generic}
\ena
to get a proper consistency equation. This ansatz is generic, and we should use for generic value of $N$. However, the solution of the self-consistency equation seems 
very difficult to be solved. We therefore, assume $N>3$ for later calculation and negrect these terms. Then we find that the remaining 
 integrations are just Gaussian integrals so that we can get a self-consistent equation for
$\Delta^\prime$ via the equation (\ref{gap}).

The gap equation of this model can be easily obtained as
\bea
(N-1)!(\frac{1}{1-\a  \Delta^\prime})^{N-2}=(N-1)N^{N-2}\Delta^\prime,
\label{relation}
\ena
which can be written as
\bea
(1-2/N)(1-3/N)\cdots(1-(N-1)/N)(\frac{1}{1-\a  \Delta^\prime})^{N-2}=\Delta^\prime.
\ena
If we can take a very "naive" planar limit, we obtain the gap equation
\bea
(\frac{1}{1-\a  \Delta^\prime})^{N-2}=\Delta^\prime.
\label{equation of deltaprime}
\ena
We should admit that this "planar" limit is too subtle. This may be originated from a simple ansatz for the correlation functions (\ref{gap}). As a diagrammatic calculation,
we should select only a small part of the diagrams to get this form of correlation functions. Therefore, we should use more
generic ansatz (\ref{generic}) to perform a more rigorous analysis. However, the gap equation becomes very complicated.  A partial
evidence of the appearance of the gap equation (\ref{equation of deltaprime}) is the following.  Instead of considering the more general ansatz, we
enhance the number of diagrams we should correct by changing the correlation functions. Namely, we enhance the number by changing the equation as
\bea
N<\prod_{k\neq
i}^{N-1}(\sum_{a_j}Q_{a_j}^k\tQ^{a_j}_k)>=(N-1)N^{N-2}\Delta^\prime\delta_{a_i}^{b_i}. 
\label{modified gap}
\ena
Since the correlation function can $<Q_i^a\tq_b^j>$ can be given by
\bea
<Q_i^a\tq_b^j>=\frac{\delta_{ij}\delta_{ab}}{1-\a \Delta^\prime},
\ena
we get a "gap equation" in the form
\bea
(\frac{1}{1-\a  \Delta^\prime})^{N-2}=(1-1/N)\Delta^\prime
\ena
As a "planar" equation, this equation leads to (\ref{equation of deltaprime}).  

In this paper, we assume the "planar" gap equation  is given by
(\ref{equation of deltaprime}). The reason we insist on the appearance of the gap equation will be given bellow. When we change variables to $\tD$
by
\bea
\frac {1}{1-\a \Delta^\prime}=1+\a  \tD,
\label {change}
\ena
we can easily find that equation (\ref {equation of deltaprime}) leads to
\bea
\tD=(1+\a\tD)^{N-1},
\ena
which is exactly the equation (\ref {valueofdtilder}). This strongly suggests that the exact planer limit of the gap equation is given by (\ref {equation of
deltaprime}). 

Let us evaluate the free energy within our approximation. Performing the Gaussian integrations, we can evaluate the free energy as
\bea
{\cal {F}}=S\ln\frac {m_1m_2\cdots m_N}{\La^N}+S (N-1)\ln (1-\a \Delta^\prime)+\a S(N-1)\Delta^\prime\frac {1}{1-\a \Delta^\prime}.
\ena
We rewrite the free energy in terms of $\tD$ by the identification (\ref{change}) to obtain
\bea
{\cal{F}}=S\ln\frac{m_1m_2\cdots m_N}{\La^N}-S\ln\tD +\a S(N-1)\tD.
\ena
This is the final form of the free energy which should be compared to the field
theoretical result(\ref{free energy of superpotential}):
\bea
{\cal{F}}=S\ln\frac{m_1m_2\cdots m_N}{\La^N}-S\ln\tD +\a S(N-2)\tD.
\ena
The difference of the last term may be related to our subtle planar limit. 
\sect{Discussions}
  We have discussed a method to obtain the effective superpotential for baryons and mesons by using the mean-field approximation. We have derived
a result which is almost identicat to the field theoretical result. However, we have assumed "planar limit" of the self-consistency equation for the
derivation. We could not prove the conjecture since we have used a simple ansatz for the correlation functions. We should use more rigorous ansatz to
get an improved result. It seems interesting that the resulting expression from the field theory is of the form which may be fully obtained by the
mean-field method. Our result may indicates the existence of some simple derivation of the superpotential for baryonic fields. 

  Another Interesting application of the mean-field method is for the gauge theories with $N_f>N_c$. In this case, the most interesting aspect is
the appearance of Seiberg duality which has been treated in the framework of the matrix models in ~\cite{Feng,FH,Ohta} However, the baryonic degree of
freedom will be required for the complete derivation of the duality. Mean-field approach may be useful to get the duality map of the fields
identification.


\end{document}